# Efficient spectral broadening and few-cycle pulse generation with multiple thin water films

Jiacheng Huang[1], Xiang Lu[1], Feilong Hu[1], Jie Long[1], Jiajun Tang[1], Lixin He[1,2], Qingbin Zhang[1,2,3], Pengfei Lan[1,2*], and Peixiang Lu [1,2]

[1]*Wuhan National Laboratory for Optoelectronics and School of Physics, Huazhong University of Science and Technology, Wuhan 430074, China*
[2]*Optical Valley Laboratory, Hubei 430074, China*
[3]*zhangqingbin@hust.edu.cn*
*\*pengfeilan@hust.edu.cn*

**Abstract:** High-energy, few-cycle laser pulses are essential for numerous applications in the fields of ultrafast optics and strong-field physics, due to their ultrafast temporal resolution and high peak intensity. In this work, different from the traditional hollow-core fibers and multiple thin solid plates, we represent the first demonstration of the octave-spanning supercontinuum broadening by utilizing multiple ultrathin liquid films (MTLFs) as the nonlinear media. The continuum covers a range from 380 to 1050 nm, corresponding to a Fourier transform limit pulse width of 2.5 fs, when 35 fs Ti:sapphire laser pulse is applied on the MTLFs. The output pulses are compressed to 3.9 fs by employing chirped mirrors. Furthermore, a continuous high-order harmonic spectrum up to the 33rd order is realized by subjecting the compressed laser pulses to interact with Kr gas. The utilization of flowing water films eliminates permanent optical damage and enables wider and stronger spectrum broadening. Therefore, this MTLFs scheme provides new solutions for the generation of highly efficient femtosecond supercontinuum and nonlinear pulse compression, with potential applications in the fields of strong-field physics and attosecond science.

## 1. Introduction

Intense ultrafast lasers with few-cycle pulse widths play an essential role in the fields of extreme nonlinear optics, strong-field physics, ultrafast optics, and attosecond science [1-7]. Such ultrafast pulses can be achieved through the nonlinear post-compression techniques, which rely on the interplay between supercontinuum generation (SCG) and compensation of dispersion [8-10].

The SCG can be achieved by self-phase modulation (SPM), self-steepening and ionization between intense laser pulses and nonlinear medium [11-14]. Extensive experimental and theoretical research has been conducted over several decades based on various nonlinear media, including gases and transparent solids [15-17]. For many years, gas-filled hollow-core fibers (HCFs) pulse compressors have gained great interest for their high beam quality and high compression factors [18,19]. However, HCFs are restricted by the incident energy and the strong nonlinear effects in the fibers due to inherent limitations on waveguide structures and the small core diameters. Pointing-stabilization setups are required for alignment and the coupling efficiency of fibers is usually low [20]. The scheme using multiple solid plates has attracted significant attention due to its simplicity, compactness, flexibility, and free-space geometric characteristics [21-26]. The generation of quasi-stationary spatial solitons in periodic-layered Kerr media (PLKM) has emerged recently as a reliable scheme to enhance the nonlinear light-matter interaction and suppress the spatial and temporal losses [27-29]. More recently, multi-pass cell (MPC) scheme has also been reported for the high energy input capacity, high efficiency (~90%), high beam quality, and spectral homogeneity across the beam profile [30-32]. The threshold of self-focusing in the solid medium is typically on the order of a few MW. Whereas in inert gas, it is much higher (~10 GW) [11]. The spectrum range of MPC may be narrower than that of MTSPs and HCFs, and the relatively low compression ratio usually requires two-stages nonlinear compression for few-cycle pulses [33-35].

Due to the higher nonlinear susceptibility and larger molecular density compared to gases [11,36], we expect physical mechanisms would be magnified in the liquid medium [37,38]. However, to our knowledge, no research has been conducted on the nonlinear compression of few-cycle pulses with liquid medium. Although supercontinuum can be obtained in condensed liquid, it is still restricted to the input energy limitation and optical loss caused by self-focusing and filamentation [39,40]. Studies show that spectrum broadening precedes pulse splitting or optical breakdown [41]. To address these challenges, we propose controlling the liquid's thickness to obtain multiple thin liquid films (MTLPs). This arrangement ensures that laser pulses solely experience spectral broadening due to self-phase modulation (SPM) in the ultrathin films. When compared to solid or gas media, the MTLFs scheme offers several potential advantages. Taking the most common water as an example, firstly, there is no permanent optical damage in the flowing water, allowing for high-intensity laser input. Liquid degradation or surface damage could be avoided by the continuous refreshment of the water [42]. Additionally, the fluidity of water ensures a fresh area for each pulse. Water is cost-effective and easily available. Its transmission efficiency of up to 99% in the visible band ensures the high-efficiency output of SCG [43]. Thirdly, the nonlinear refractive index of water ($5.7 \times 10^{-20}$ m$^2$/W) is much higher than fused silica (FS) [11,44], theoretically leading SCG with wider spectral range and stronger intensity. Based on the above consideration, multiple ultrathin water films show promise as a better spectral broadening scheme, replacing the traditional gas-filled HCFs and solid thin plates.

In this paper, we demonstrate the generation of sub-4-fs few-cycle laser pulses by octave-spanning supercontinuum based on the multiple thin water films (MTLFs) scheme. The obtained supercontinuum covers from 380 to 1050 nm with an efficiency of up to 88.7%. After dispersion compensation by chirped mirrors, the chirped pulses are compressed from 35 fs to 3.9 fs successfully. The high quality of the compressed pulses is further confirmed by high-harmonic generation. A continuous high-order harmonic spectrum up to the 33rd order is achieved based on the compressed few-cycle laser pulses.

## 2. Nonlinear compression experiment

### 2.1 The generation of supercontinuum

Firstly, we investigate the SCG using multiple thin liquid films (MTLFs). The schematic of our experimental set-up is shown in Fig. 1. A commercial 800 nm Ti:sapphire laser system (Astrella-USP-1K, Coherent, Inc.) is employed as the pump source, which delivers 35 fs pulses at a repetition rate of 1 kHz. An iris is used to reduce the beam size to 6 mm for a larger Rayleigh range at the focus. Through the convex lens of f = 2000 mm, the Rayleigh range is measured to be 16.1 cm with the peak power of focused beam at ~17 GW (pulse energy 0.6 mJ). Multiple ultrathin wire-guided free-flowing water films are employed in our experiment, as depicted in Fig. 1(b). Benefiting from the balance between high surface tension and gravity of the water, a flow guided by two Ni-80 wires with a diameter of 200 μm forms a thin water film [44].

Laser pulses would only experience spectral broadening due to SPM, and exit the film before filamentation and optical damage occur. After passing through the film, the beam gets focused in the air and then diverges. The divergent beam would be refocused by the self-focusing effect inside the next film. This method can circumvent the filamentation and optical breakdown, allowing for the enhancement of the nonlinear Kerr interaction inside the liquid.

By carefully optimizing and adjusting the positions of films near the focus, we achieve the optimal conditions for SCG with three water films. The first water film is placed ~100 mm in front of the focal point, and the spacings between the films are 118 mm and 50 mm, respectively. It is worth noting that the thickness of the water films can be easily adjusted by controlling the water flow rate and the spacing between two wires, which will provide flexibility in tuning the nonlinear phase shift accumulated in the water films. To measure the nonlinear broadening spectrum, we utilize a CCD optical spectrometer (AvaSpec-2048FT-SPU) and position it 60 cm downstream from the focus.

The far-field beam profile after spectrum broadening is captured on a flat piece of white paper and imaged onto a camera, as depicted in Fig. 1(c). The beam spot represents a high-quality transverse mode with zeroth-order Bessel function, which is similar to the phenomenon observed in the scheme of FS plates [11]. In general, the white light beam generated during filament in gases and condensed media would consist of bright round central zone surrounded by conical emission [45]. An iris aperture is employed to block outer diffraction rings before recollimation. The energy proportion of the center spot is 79.7%, much higher than that of in HCFs systems (~50%) [19]. The beam instability caused by the generation of water films can be further optimized by improving the water film workpiece.

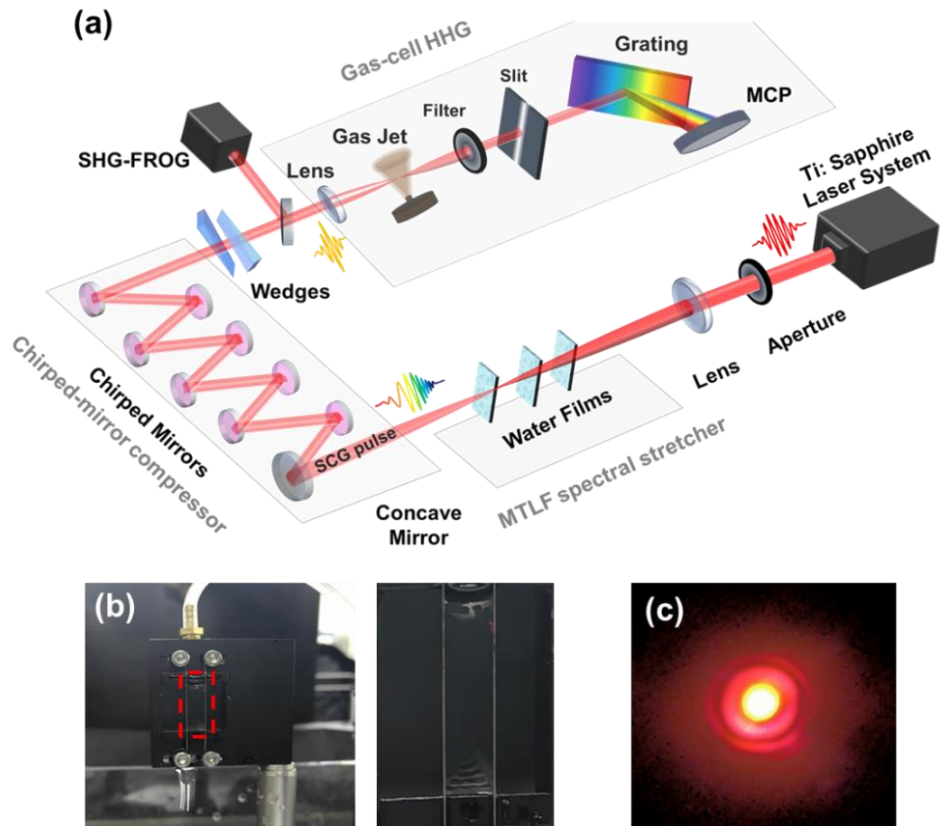

**Fig. 1** (a) Schematic of the NLO compression and HHG experimental setup. MCP, microchannel plate. (b) The gravity-driven ultrathin water film workpiece. (c) The beam profile is collected on flat paper and imaged on a camera.

The spectrum evolution of fundamental femtosecond pulses after the MTLFs is depicted in Fig. 2(a). The pulse spectrum broadens incrementally after passing through each water film. After the first water film, the spectrum expands from its initial range of 740-860 nm to 580-960 nm. As the second film is introduced, further spectral broadening occurs, resulting in a range of 480-985 nm. In the first two films, the broadened spectra exhibit symmetrical expansion on both the red and blue sides, which is in accordance with the process of SPM [46]. However, the more pronounced blue extension is observed after propagating through the last water film due to the self-steepening and ionization effects, as demonstrated previously in the FS plates broadening scheme [22]. Three water films are employed for maximum broadening, no significant further spectral broadening effect is observed when adding more water films. Ultimately, at an intensity level of -30 dB, the final broadened spectrum covers a range of 380-1050 nm, corresponding to the Fourier Transform Limit (FTL) pulse duration of 2.5 fs. The supercontinuum output energy reaches 0.52 mJ with an efficiency of 86.7%. The total energy loss of ~10% through the three water films can be attributed to multiphoton absorption and ionization in water [47]. Additionally, diffuse reflection caused by the non-smooth surface of water films also contributes to certain energy loss. Note that the critical self-focusing threshold power $P_{crit}$ of water is estimated to be several MW [48]. Here, $P_{crit} = \lambda_2/2\pi n_0 n_2$, $n_0$ is the linear refractive index and $n_2$ represents the nonlinear refractive index [11]. In our experiment, due to the use of thin liquid films, laser pulses with a peak power of ~17 GW can be employed, significantly exceeding the critical power $P_{crit}$ of water.

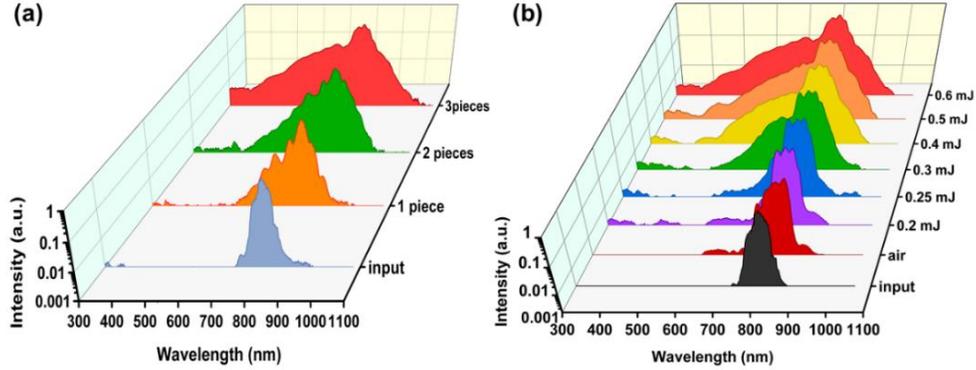

Fig. 2 (a) Experimentally measured spectra after each water film at the incident energy of 0.6 mJ. (b) Broadened spectra with 3 pieces of water films under different incident pulse energies.

The applicable range of input laser energy in our MTLFs scheme is also characterized. By utilizing a combination of a half-waveplate (HWP) and a polarizer, we continuously reduce the incident pulse energy from 0.6 to 0.2 mJ at 0.1 mJ intervals. The input beam waist size and the position of water films remain unchanged. Thus, the incident intensity on the water films for SCG solely depends on the incident pulse energy. The obtained broadening spectra under different input energies are shown in Fig. 2(b). As the incident laser energy decreases, the range of broadened spectrum reduces significantly. When the input energy drops below 0.2 mJ, no considerable spectral broadening occurs. It indicates that the nonlinear intensity is too weak to achieve SPM in the water films under this circumstance. In the case of low energy laser input, the focusing lens with a smaller focal length can be employed to obtain sufficiently high intensity for SCG.

Additionally, we conduct the supercontinuum generation with MTLFs under high energy laser input. 1 mJ energy and 1 kHz repetition rate laser pulses are injected into our thin water films. In this case, the intense average power will lead to the evaporation of the thin water film and the filamentation in water under the conditions of the previous experimental setup. To address this issue, adjustments are made to optimize supercontinuum generation by modifying the positions and spacings between the water films. Compared to the relatively low intensity input, the spectral broadening through air is more pronounced with 1 mJ energy input. As depicted in Fig. 2(b), the final spectral broadening covers a range from 420 to 1000 nm at the -30 dB bandwidth. Despite the difference in incident energies, the spectral configurations are similar in both cases. Furthermore, after passing through three water films, the output energy is measured to be 0.84 mJ with a conversion efficiency of 84%. As for much higher pulse energy input, a focusing lens with larger focal length is required. It is necessary to maintain relatively constant nonlinear intensity on the water films for effective SCG. The large range of permissible pulse energies input of our MTLFs setup makes it a promising platform for various SCG and nonlinear compression applications.

For comparison, we also conduct a contrasting experiment using traditional thin fused silica (FS) plates as the nonlinear media for SCG. Under similar laser input conditions, the supercontinuum in the multiple FS thin plates scheme expands from 450 to 980 nm. The total thickness of the FS plates is 600 μm. As depicted in Fig. 3, it is evident that the spectrum achieved with the MTLFs setup exhibits wider range and higher intensity, indicating the stronger nonlinear interaction in the water films. Moreover, the intensity of the broadened spectrum below 500 nm based on water films is even ten times higher than that achieved with FS plates. These results not only validate the feasibility of utilizing ultrathin water films for SCG, but also highlight their superiority over traditional FS plates in terms of spectral broadening and intensity enhancement, showing great potential for subsequent nonlinear pulse compression.

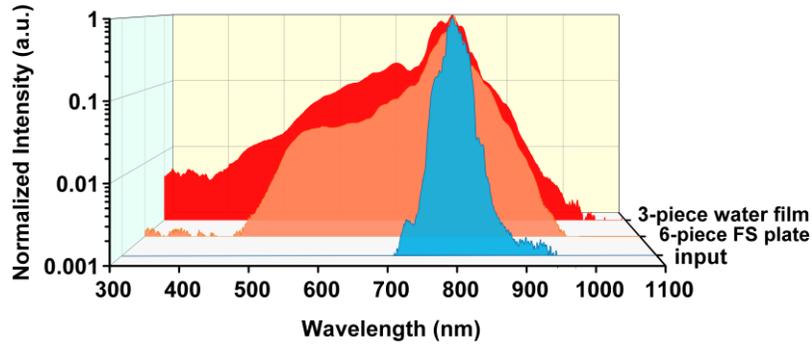

**Fig. 3** The comparison of broadened spectra between MTLFs and FS plates scheme.

To gain further insight into the experimental results of SCG, we conduct numerical solutions of the generalized nonlinear Schrodinger equation (NLSE) to investigate the pulse propagation [49] (see Supplement). The simulated spectral configuration closely agrees with the experimental results. The Kerr nonlinear optical response, including SPM and the self-steepening, acts jointly with the dispersion effect to stretch the spectrum. The Raman scattering and ionization effect are also considered. The influence of the nonlinear refractive index $n_2$ of the medium on the spectral broadening is verified theoretically. The wider spectrum obtained in our MTLFs scheme is greatly affected by the larger nonlinear refractive index $n_2$ of water compared to fused silica. These theoretical results indicate that the compressed pulse duration would be shorter under relatively large nonlinear refractive index $n_2$ and low group velocity dispersion $\beta_2$, which is in agreement with the Ref. [51].

## 2.2 The pulse compression

To compress the pulse duration, four pairs of chirped mirrors (Ultrafast Innovations) with 450-1000 nm bandwidth and GDD = -40 $fs^2$ are utilized to compensate for the chirp of the output pulses. Additionally, a pair of wedges is incorporated to finely tune the dispersion. After the compression elements, the laser is focused using a focal lens of f = 150 mm. The efficiency of the output pulse energy, measured after the chirped mirrors, is approximately 91.4%. Moreover, the center spot of output beam contains energy of ~0.37 mJ.

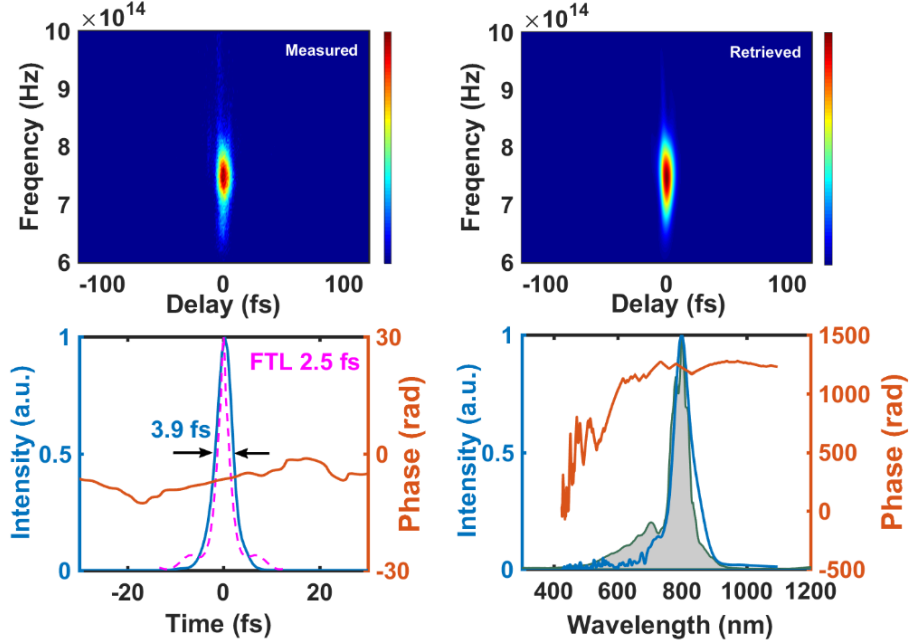

**Fig. 4** Characterization of the 3.9 fs few-cycle pulses compressed from the supercontinuum by homemade SHG-FROG. (a), (b) Measured and retrieved FROG traces. (c) Retrieved pulse envelop (solid blue line), FTL pulse of 2.5 fs (dashed line), and temporal phase (solid orange line). (d) Retrieved spectrum (solid blue line), the spectrum taken by the CCD spectrometer (shaded region), and the spectral phase (solid orange line).

A homemade scanning second-harmonic generation frequency-resolved optical gating (SHG-FROG) measurement is utilized to record the temporal and spectral characterizations of the compressed pulses. Figure 4(a) and (b) represent the measured and retrieved FROG traces of the compressed pulses, respectively, with the retrieved error of 0.005. Figure 4(c) illustrates the reconstructed temporal profile of pulse duration (blue solid) and phase (orange solid). The duration of pulses is measured to be 3.9 fs by FROG, corresponding to 1.6 FTL (FTL = 2.5 fs) of the measured spectrum. And the pulse compression factor $K_c$ is approximately 10, which is comparable to the reported highest compression factor using a single compression stage [52]. The difference in pulses characteristics may be attributed to the bandwidth limitations of the chirped mirrors, as some spectral components are not fully compensated for dispersion. As shown in Fig. 4(d), the retrieved spectrum in blue line agrees well with the actual spectrum measured by the spectrometer (shaded area). The spectral phase is relatively flat in the spectral range of 600-1000 nm, suggesting the effect of SPM plays a dominant role in this spectral broadening region and dispersion has been effectively compensated in the experiment. However, below 600 nm, the spectral phase exhibits modulation due to possible contributions from the self-steepening effect and ionization.

### 3. HHG experiment

To further demonstrate the quality of the compressed pulses, we carry out the high-order harmonic generation experiment by interacting with Kr gas after compressing the supercontinuum. After the chirped mirrors compressor stage, certain loss is induced by multiple reflections on the silver mirrors and the vacuum window. The compressed few-cycle pulses laser with the energy of 0.26 mJ is focused onto a gas target inside a vacuum chamber using a concave mirror with a focal length of 150 mm. The generated high-order harmonics are detected by a homemade flat-field soft x-ray spectrometer, which consists of a 0.2 mm-wide, 15 mm-height entrance slit, a flat-field grating (1200 grooves mm[-1]), and a microchannel plate (MCP) backed with a phosphor screen [53]. The spectrally resolved images are recorded using a charge-coupled device (CCD) camera.

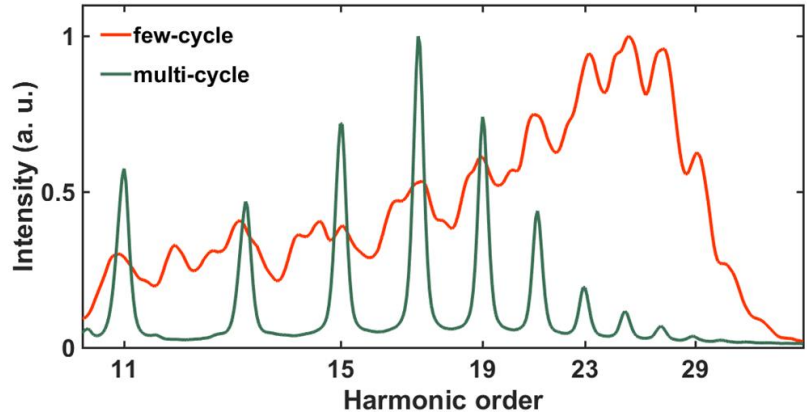

**Fig. 5** Experimentally measured high-order harmonic spectra generated in krypton driven by compressed few-cycle pulses (red) and uncompressed multi-cycle pulses (green).

As shown in Fig. 5, the cut-off region of the high-order harmonic spectrum is continuous and smooth. And the signal of the 33rd harmonic with photon energy up to 51 eV is observed in the generated supercontinuum, corresponding to an approximate peak intensity of $1.8\times10^{14}$ W/cm$^2$ in the interaction region. A certain amount of dispersion would be introduced during propagation through long-distance air and the vacuum chamber window. In contrast to the compressed few-cycle pulses, we also inject the uncompressed pulses (35 fs in the full-width-at half-maximum (FWHM)) into the chamber. The corresponding odd harmonics appear discrete in the spectrum and the peaks are noticeably sharper. The observed supercontinuum signal in the 33rd harmonic demonstrates the potential for the follow-up attosecond pulse generation and attosecond science applications.

## 4. Conclusion

In this work, we present the first demonstration of octave-spanning SCG based on multiple ultrathin water films, covering a wide range from 380 to 1050 nm with an efficiency of 86.7%. Few-cycle 3.9 fs laser pulses are achieved successfully after dispersion compensation by chirped mirrors. The high quality of the compressed pulses is further confirmed by the generation of a continuous high-harmonic spectrum up to the 33rd order. This MTLF scheme provides valuable solutions for SCG covering from ultraviolet to near-infrared and nonlinear pulse compression with few-cycle durations. Additionally, liquids with larger nonlinear refractive indices can be employed as the nonlinear media to achieve an even broader spectrum, such as certain alcohol. Furthermore, this work has potential applications in the fields of high-order harmonic generation and attosecond pulse generation, contributing to the advancement of ultrafast science and strong-field physics.

**Funding.** National Nature Science Foundation of China (No. 91950202, No. 12225406, No.12074136, No.12021004).

**Acknowledgments.** We thank Xicheng Zhang, and Yiwen E for the helpful discussions on the thin liquid films. We gratefully acknowledge the financial support from the National Nature Science Foundation of China.

**Disclosures.** The authors declare no conflicts of interest.

**Data availability.** Data underlying the results presented in this paper are not publicly available at this time but may be obtained from the authors upon reasonable request.